\documentclass[a4paper,10pt]{article}
\usepackage[utf8]{inputenc}
\usepackage[dvipsone]{graphicx}
\usepackage{amssymb}
\usepackage{textcomp}
\usepackage{amsmath}
\usepackage{floatflt}
\usepackage[top=2cm, bottom=2.5cm, left=1in, right=1in]{geometry}
\title{The evidence of quasi-free positronium state in GiPS-AMOC spectra of glycerol}
\author{D. Zvezhinskiy$^{1,*}$ \and M. Butterling$^{2}$ \and A. Wagner$^{2}$ \and R. Krause-Rehberg$^{3}$ \and S.V. Stepanov$^{1}$}

\begin{document}
\date{}
\maketitle

\begin{abstract}
We present the results of processing of Age-Momentum Correlation (AMOC) spectra that were measured for glycerol by the Gamma-induced positron spectroscopy (GiPS) facility. Our research has shown that the shape of experimental $s(t)$ curve cannot be explained without introduction of the intermediate state of positronium (Ps), called quasi-free Ps. This state yields the wide Doppler line near zero lifetimes. We discuss the possible properties of this intermediate Ps state from the viewpoint of developed model. The amount of annihilation events produced by quasi-free Ps is estimated to be less than 5\% of total annihilations.  In the proposed model, quasi-free Ps serves as a precursor for trapped Ps of para- and ortho-states. 
\end{abstract}


{\let\thefootnote\relax\footnotetext{$*$ corresponding author, zmitja@yandex.ru\\
${}^{1}$ Institute for Theoretical and Experimental Physics, B. Cheremushkinskaya, 25,
117218, Moscow, Russia\\
${}^{2}$ Institute of Radiation Physics, Helmholtz-Zentrum, Dresden-Rossendorf, P.O. Box
510119, 01314, Dresden, Germany\\
${}^{3}$ Martin-Luther-University Halle-Wittenberg, Dept. of Physics, 06099, Halle,  Germany}}

\section{Introduction}
Among the set of positron annihilation methods, the Age-Momentum Correlation (AMOC) technique yields the most complementary information about the fate of positron inside the investigated medium, that is its lifetime (or age) and the energy of annihilation gamma-quantum. 
Although these properties may tell researcher more information together than their separate measurements (i.e. a lifetime and Doppler spectra respectively), the task of interpretation of 2-dimensional spectrum is practically more difficult. 
Recently we applied an approach based on such 2-dimensional interpretation of AMOC spectrum, that helped us to verify the diffusion-recombination model of Ps for the water which spectrum has been measured with Gamma-induced modification of AMOC technique where slowing-down radiation gives birth to positron \cite{jop13}. 

The lack of 2-dimensional approach is a relatively low count rate for each bin of the spectrum, that results in a large value for a relative error estimation. Additionally, the significant amount of bins in AMOC spectrum have no events, that makes it not reasonable to apply the chi-square approach for fitting procedure.  

To verify the Ps formation mechanism by means of interpretation of the GiPS-AMOC spectra, we propose an alternative approach, which eliminates some of these issues. It has been applied to a glycerol at the set of temperatures between 46 and 122$^{\rm o}$C. 
Glycerol has been chosen as a convenient modeling substance because of its physical properties that significantly change with temperature, namely viscosity, which is of greater interest for the current experiment, as it should influence the estimated time of medium reorganization in the presence of Ps and the growh of its bubble state. 

\section{Experimental details}
The general description and advantages of GiPS-AMOC technique and its comparison with conventional technique were discussed previously \cite{Butt11}. We consider it mainly because of a perfect signal-to-background ratio and the absence of dissolved or sandwiched positron source.  Before the experiment, the amount of glycerol (99.95\% grade) was heated in a 
vacuum oven in order to remove water before measurement. During the measurement the nitrogen has been bubbled through the 
glycerol and the entry of new air from outside has been prevented. The supply system located inside the irradiation chamber and exposed to bremsstrahlung (which gives birth to positrons inside the investigated volume) consisted of in-chamber kapton vessel, which was connected with plastic tubes to a pump and heating systems, located outside the chamber. This scheme eliminated the appearance of positrons inside the most of auxiliary systems. 

The processing of experimental spectra has been done as follows.
From the AMOC spectrum, we extracted three experimental curves:

\textbf{1.} The lifetime spectrum $N(t)$, which is calculated as a projection of AMOC spectrum on time axis. For each $t$, the number of events was summed for events with energies of $E = mc^2 \pm$ 5 keV.

\textbf{2.} The Doppler spectrum $D(E)$, which is a projection of AMOC spectrum on energy axis. To obtain this dependence for each energy, the number of events was summed over all ages in the given spectrum.

\textbf{3.} The ``sharpness'' curve, or $s(t)$ curve, which is calculated at a given $t$ as a ratio of two sums of events, namely the number of events in a ``narrow'' range ($E = mc^2 \pm$ 0.7 keV) and the number of events in a ``full'' range (chosen as $E = mc^2 \pm$ 3 keV).

As a measure of goodness-of-fit, the collective chi-square functional was chosen, which is introduced as a sum of squared residuals between models (see below) and corresponding experimental data, weighted with inverse error estimates. Its minimization yielded the results reported below.


\section{Gauss-exponential model with quasi-free Ps}

The main purpose of newly developed theoretical model is to describe the peculiarity of the data, particularly of $s(t)$ curve, where a peak in the region of ``zero channel''\footnote{It is a particular bin at time axis where positron lifetime is equal to zero} is observed for all temperatures of glycerol. 

This feature means that at short lifetimes the shape of Doppler spectrum varies from wide to narrow and then back to wide. We propose the following explanation to this phenomenon. The hot positron entering the investigated medium quickly loses its energy (this time is estimated as some picoseconds for a broad range of liquids), and then it becomes solvated by the surrounding medium at a time of about 10 ps (estimates for electron trapping, see \cite{glyrad}). 
However, some part of these primary positrons form bounded state, which is called positronium (Ps). Because of electroneutrality, Ps interacts weaker with medium staying untrapped for some time. If such delocalized electron-positron pair is observed via its annihilation, until its trapping occurs in the medium, its parameters of annihilation should not significantly differ from those of free (untrapped) e+. Moreover, Ps should find a stuctural defect with suitable size to be trapped, and our estimats give the necessary radius of 2 \AA\ for this preexisting cavity. This search may take some time, which is expected to extract from the experiment. 
Thus, in the framework of the proposed model, the appearance of a wide component in Doppler spectrum at short lifetimes is explained as annihilation of delocalized (or quasi-free) Ps. 

\begin{floatingfigure}[l]{55mm}
 \centering
 \includegraphics[scale=0.55]{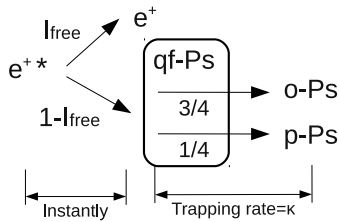}
 \caption{The model of positron formation with quasi-free state}
 \label{fig:trapscheme}
\end{floatingfigure}

Our current model is mainly similar to the one used in \cite{Schneider}, especially in the part of physical origins of ``juvenile'' broadening and general way of model construction, except some details which will be highlighted below.

Let us show how quasi-free Ps is introduced to the theoretical model.
After the process of ionizational slowing down, positron either remains in a free state, or it becomes binded with any of intratrack electrons and forms quasi-free positronium state (qf-Ps). Then quasi-free state is trapped with the rate of $\kappa$ to localized p-Ps and o-Ps states, see Fig. \ref{fig:trapscheme}.
In the proposed model we assume that this quasi-free Ps state possesses with a Doppler line similar to free e$^+$ state, and its lifetime corresponds to $1/\lambda_{e+}$. 

The formulated model yields the following set of kinetic equations for e+ and Ps concentrations.
\begin{align}
  \frac{d c_{qf}(t)}{dt} &=-\kappa \cdot c_{qf} - \lambda_{e+} \cdot c_{qf};\ \ c_{qf}(0)=1-I_{free};\\
  \frac{d c_{e+}(t)}{dt} &=-\lambda_{e+} \cdot  c_{e+};\ \ c_{e+}(0)=I_{free}; \\
  \frac{d c_{\rm{p-Ps}}(t)}{dt} &= 0.25 \cdot \kappa \cdot c_{qf} - \lambda_{\rm{p-Ps}} \cdot c_{\rm{p-Ps}};\ \ c_{\rm{p-Ps}}(0)=0; \label{eq:7}\\
  \frac{d c_{\rm{o-Ps}}(t)}{dt} &= 0.75 \cdot \kappa \cdot c_{qf} - \lambda_{\rm{o-Ps}} \cdot c_{\rm{o-Ps}};\ \ c_{\rm{o-Ps}}(0)=0. \label{eq:8}
\end{align}
Here, $I_{free}$ is the fraction of positrons that does not form Ps after the ionizational stopping.
This set of equations has the following solutions:
\begin{align}
 c_{qf}(t) &= (1-I_{free})\cdot \exp{\left(-(\kappa+\lambda_{e+}) t\right)};\label{eq:sol1}\\
 c_{e+}(t) &= I_{free}\cdot \exp{\left(-\lambda_{e+} t\right)};\\
 c_{\rm{p-Ps}}(t)&=(1-I_{free})\cdot 0.25\cdot \frac{\left[\exp{\left(-\lambda_{\rm{p-Ps}}\ t\right)}-\exp{\left(-(\kappa+\lambda_{e+})\ t\right)}\right]}{1+(\lambda_{e+}-\lambda_{\rm{p-Ps}})/\kappa} ;\\
 c_{\rm{o-Ps}}(t)&=(1-I_{free})\cdot 0.75\cdot \frac{\left[\exp{\left(-\lambda_{\rm{o-Ps}}\ t\right)}-\exp{\left(-(\kappa+\lambda_{e+})\ t\right)}\right]}{1+(\lambda_{e+}-\lambda_{\rm{o-Ps}})/\kappa}.\label{eq:sol4}
\end{align}
To obtain a lifetime spectrum 
from here, we sum items from eq (\ref{eq:sol1})-(\ref{eq:sol4}) multiplied by corresponding annihilation rates:
\begin{equation}\label{eq:NT}
 N_{\rm theor}(t) \simeq c_{qf}\ \lambda_{e+} + c_{e+}\ \lambda_{e+} + c_{\rm{p-Ps}}\ \lambda_{\rm{p-Ps}} + c_{\rm{o-Ps}}\ \lambda_{\rm{o-Ps}}
\end{equation}
As we should take into account the finite time resolution, we replace the normalized exponential decay probability density function ($\lambda \cdot \exp{(-\lambda\cdot t)}$) with its convolution with Gaussian resolution curve, $G(t)=\exp\left(-\frac{(t-t_0)^2}{2 \sigma^2}\right)/\sqrt{2 \pi \sigma^2}$, which has the following closed form \cite{ryzhik}:
$
f(t)=\lambda/2 \cdot \exp{\left(-t^2 / (2\sigma^2)\right)}\cdot
       \exp{\left(\beta\ \gamma^2\right)}\cdot\left(1-{\rm Erf}(\gamma\ \sqrt{\beta})\right).
$ Here 
$\beta=\sigma^2/2$, $\gamma=\lambda-t/\sigma^2$, $\sigma$ is the parameter (standard deviation) of resolution curve, ${\rm Erf}$ denotes the error function.

We used the following relation to evaluate the shape of $s(t)$ curve from given binned AMOC experimental data $C(E_i,t_j)$:
\begin{equation}
  s_{\rm exp}(t_j) =\frac{\Sigma_{E_i=mc^2-0.7\mbox{\begin{scriptsize}keV                                             \end{scriptsize}}}^{E_i=mc^2+0.7\mbox{\begin{scriptsize}keV                                                                                                                                                     \end{scriptsize}}}\,C(E_i,t_j)\cdot(1-f_{\rm{bg},1})}{\Sigma_{E_i=mc^2-3\mbox{\begin{scriptsize}keV                                                                                                                                                                                                                                                                                                                                                                                                             \end{scriptsize}}}^{E_i=mc^2+3\mbox{\begin{scriptsize}keV                                                                                                                                                                                                                                                                                                                                                                                                                                                                                                                                                                                                                                                                                                                                                                                                                                                                                   \end{scriptsize}}}\,C(E_i,t_j)\cdot(1-f_{\rm{bg},2})}
\end{equation}
Here we take into account a total fraction of background events $f_{\rm{bg},1..2}$, which are produced by scattered gamma-quanta and a Compton effect inside the energy detectors. This fraction is obtained by means of the following procedure.
At first, we fit Doppler spectrum with Gaussian components and a stepwise curve as an additive background. After fitting we know the total proportion of background and corresponding parameters of background component. Then we calculate the corrections for numerator and denominator and apply them for each time $t_i$ presuming that certain slice of AMOC spectrum along time axis contains the same ratio of background events as the integral Doppler spectrum.

The equation for theoretical model for $s(t)$  is derived from eq. (\ref{eq:NT}) by taking into account, that every its item should be multiplied by the integral of Doppler line with Gaussian shape, taken in the range of $mc^2 \pm$ 0.7 keV for numerator and $mc^2 \pm$ 3 keV for denominator:
\begin{equation}\label{eq:st}
s_{\rm theor}(t)=\frac{(c_{qf}+c_{e+})\lambda_{e+}H_1(\sigma_{e+}) + c_{\rm{p-Ps}}\lambda_{\rm{p-Ps}}H_1(\sigma_{\rm{p-Ps}}) + c_{\rm{o-Ps}}\lambda_{\rm{o-Ps}}H_1(\sigma_{\rm{o-Ps}})}
{(c_{qf}+c_{e+})\lambda_{e+}H_2(\sigma_{e+}) + c_{\rm{p-Ps}}\lambda_{\rm{p-Ps}}H_2(\sigma_{\rm{p-Ps}}) + c_{\rm{o-Ps}}\lambda_{\rm{o-Ps}}H_2(\sigma_{\rm{o-Ps}})}.
\end{equation}
Here $H_i(\sigma)$ is an integral of Doppler component over E in the range of ($mc^2-\Delta_i$ ; $mc^2+\Delta_i$), where $\Delta_1$=0.7keV, and $\Delta_2$=3keV. They are calculated as
$
 H_i(\sigma)=\int_{mc^2-\Delta_i}^{mc^2+\Delta_i} \exp{\left(- \frac{ (E-mc^2)^2}{2\sigma^2} \right)} / (\sqrt{2\ \pi}\sigma) dE = {\rm Erf}\left(\Delta_i / (\sqrt{2}\ \sigma)\right).
$

To construct the model for Doppler spectrum, we need to calculate the contribution of corresponding Gaussian components for ``free'' positrons, ortho- and para-positronium. They are given by the following formulae, which are derived from (\ref{eq:sol1})-(\ref{eq:sol4}):
\begin{align}
h_{free}&=(1-I_{free})\frac{\lambda_{e+}}{\lambda_{e+}+\kappa}+I_{free}; \label{eq:h1}\\
h_{\rm{p-Ps}}&=(1-I_{free})\frac{0.25\ \lambda_{\rm{p-Ps}}}{1+(\lambda_{e+}-\lambda_{\rm{p-Ps}})/\kappa}\cdot(1/\lambda_{\rm{p-Ps}}-1/(\kappa+\lambda_{e+}));\\
h_{\rm{o-Ps}}&=(1-Ifree)\frac{0.75\ \lambda_{\rm{o-Ps}}}{1+(\lambda_{e+}-\lambda_{\rm{o-Ps}})/\kappa}\cdot(1/\lambda_{\rm{o-Ps}}-1/(\kappa+\lambda_{e+})). \label{eq:h3}
\end{align}
Here we took into acctount that the width of Doppler line for free e+ and qf-Ps are introduced as equal.
Now we obtain the expression for Doppler spectrum:
\begin{equation}\label{eq:DE}
 D_{\rm theor}(E)=\sum_{i=1}^3  \frac{h_i}{\sqrt{2\ \pi}\sigma_i}\exp{\left(\frac{ (E-mc^2)^2}{2\sigma_i^2} \right)} + B(E).
\end{equation}
The background component $B(E)$ has been previously fitted to processed spectrum with a stepwise function.

The resolution function for energy detector is taken into account by presuming its Gaussian shape, and, given that the shape of Doppler components is the same, the standard deviation which enters eq. (\ref{eq:DE}) should be calculated as:
$
 \sigma_i=\sqrt{\sigma_{rf}^2+\sigma_{i,unc}^2}.
$
The parameter of resolution function $\sigma_{rf}$ is found from independent experiments and $\sigma_{i,unc}$ is uncorrected standard deviation for $e+$, p-Ps and o-Ps Doppler components.

\section{Results}
The collective fit of three theoretical models (\ref{eq:NT}), (\ref{eq:DE}), (\ref{eq:st}) to corresponding experimental data resulted in a set of obtained model parameters, that are discussed in this section.
As a reference data for comparison with new results, we used a 3-exponential decomposition of lifetime spectra measured by Duplatre and reported in \cite{gly2009}. 


Annihilation rates for para- and ortho-positronium were parametrised via a contact density parameter $\eta(T)$ (e.g. see \cite{ste11}), that  was not fixed because we were unable to find its value for glycerol in the availiable literature. 
The obtained values of $\eta$ are about 0.6(2) without having any noticeable temperature trend.
\begin{floatingfigure}[l]{75mm}
 \centering
 \includegraphics[scale=0.45]{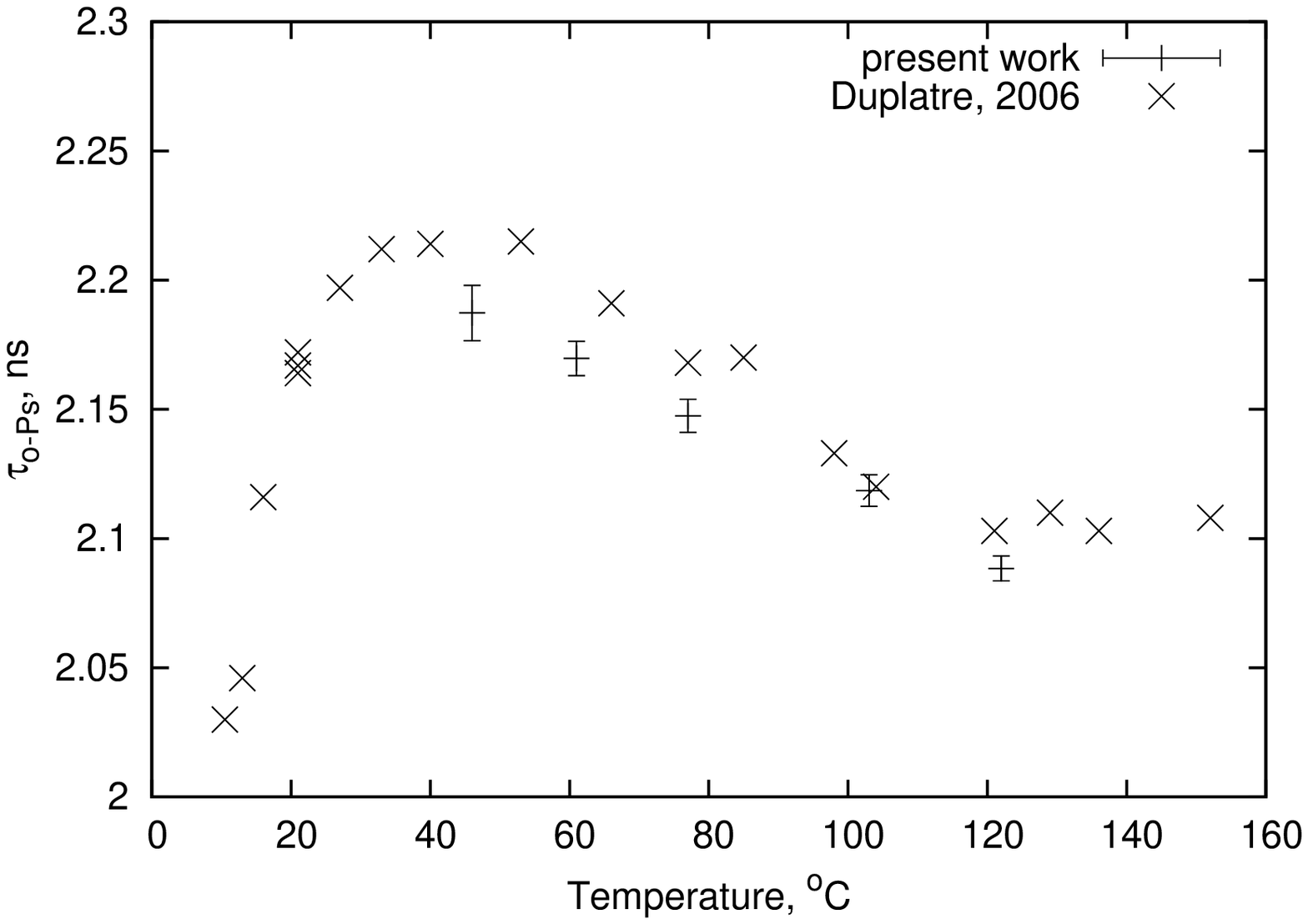}
 \caption{Temperature dependence of $\tau_{\rm o-Ps}$ for glycerol}
 \label{fig:Tpps}
  \includegraphics[scale=0.45]{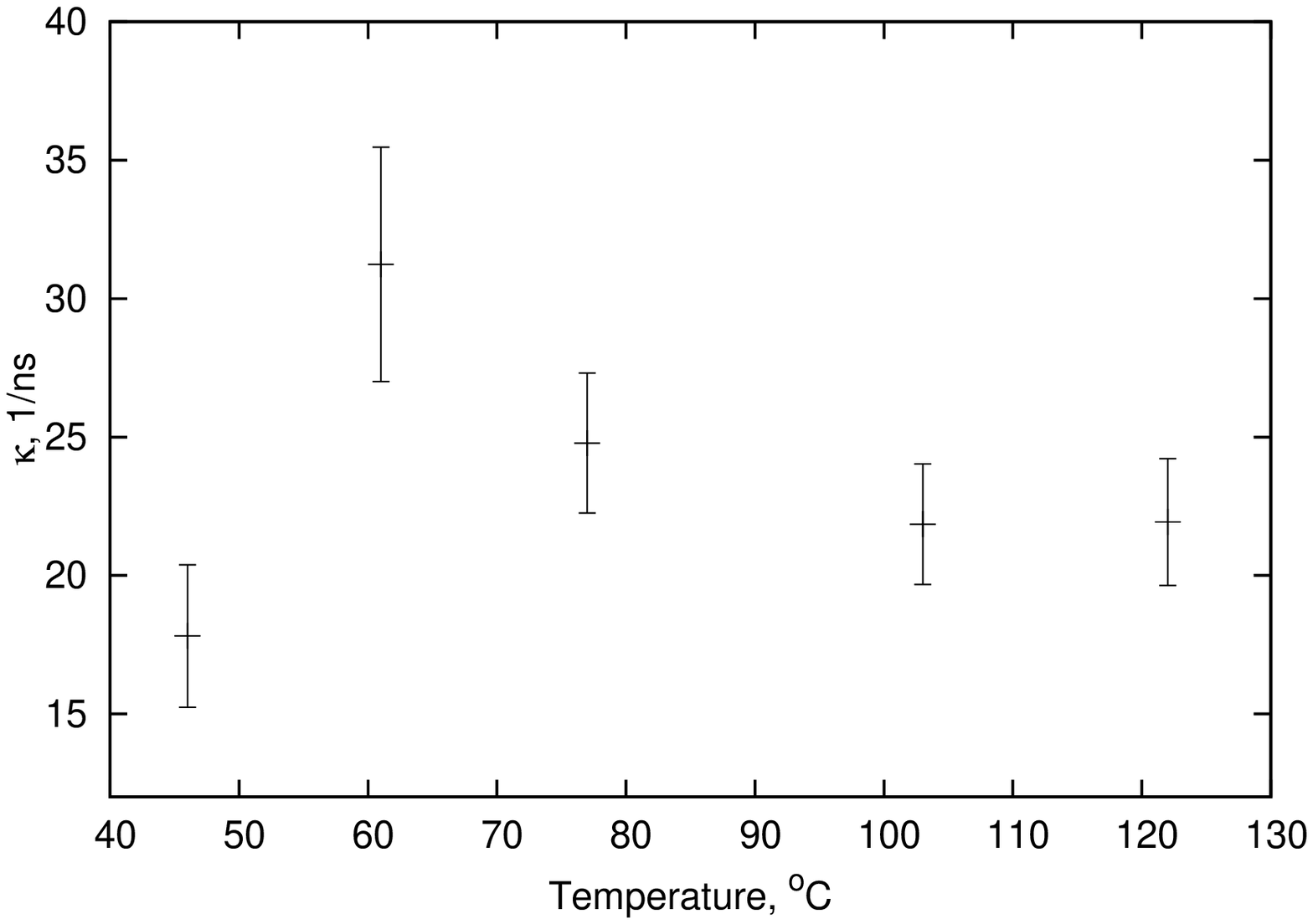}
 \caption{Temperature dependence of $\kappa$ for glycerol}
 \label{fig:kap}
\end{floatingfigure}
The lifetime of free e+ and qf-Ps is found to be 0.3(7) ns through all measured temperatures. The dependency for lifetimes of ortho-positronium on temperature is drawn on Fig. \ref{fig:Tpps}.
It is in a good agreement with the reference data. The differences between them may be caused by fixing the formation ratio of o-Ps and p-Ps in eq. (\ref{eq:7}), (\ref{eq:8}) to 3/1 in the present work, whereas it was obtained as a free parameter from the lifetime data. The lifetime of o-Ps shows the falling trend which is similar to the previously reported results. This behaviour is discussed in \cite{gly2009} and it is caused by the neglecting of Ps chemical reactions in the model.  


Standard deviations $\sigma_i$ for the shape of Doppler line is equal to 1.1(6) keV for free e+ and 0.9(8) keV for o-Ps. 
They don't show any temperature trend as well. The values for $\sigma_{\rm p-Ps}$ are varying from 0.3 keV (46$^{\rm o}$C) up to 0.2 keV (122$^{\rm o}$C).

The trapping rate $\kappa$ for quasi-free Ps turns out to be in the range of 17 ... 30 1/ns (Fig. \ref{fig:kap}), which corresponds to the time of 30 ... 60 ps. Based on estimations proposed by Stepanov \cite{stebubble}, the time of Ps bubble growth varies from 6 ns at 50$^{\rm o}$C to 1 ns at 70$^{\rm o}$C. 
 This means that the limiting stage for Ps trapping is the search of an ``empty'' space in the medium, which should be large enough to allow the trapping of Ps. 

\begin{figure}[h!]
 \centering
 \includegraphics[scale=0.43]{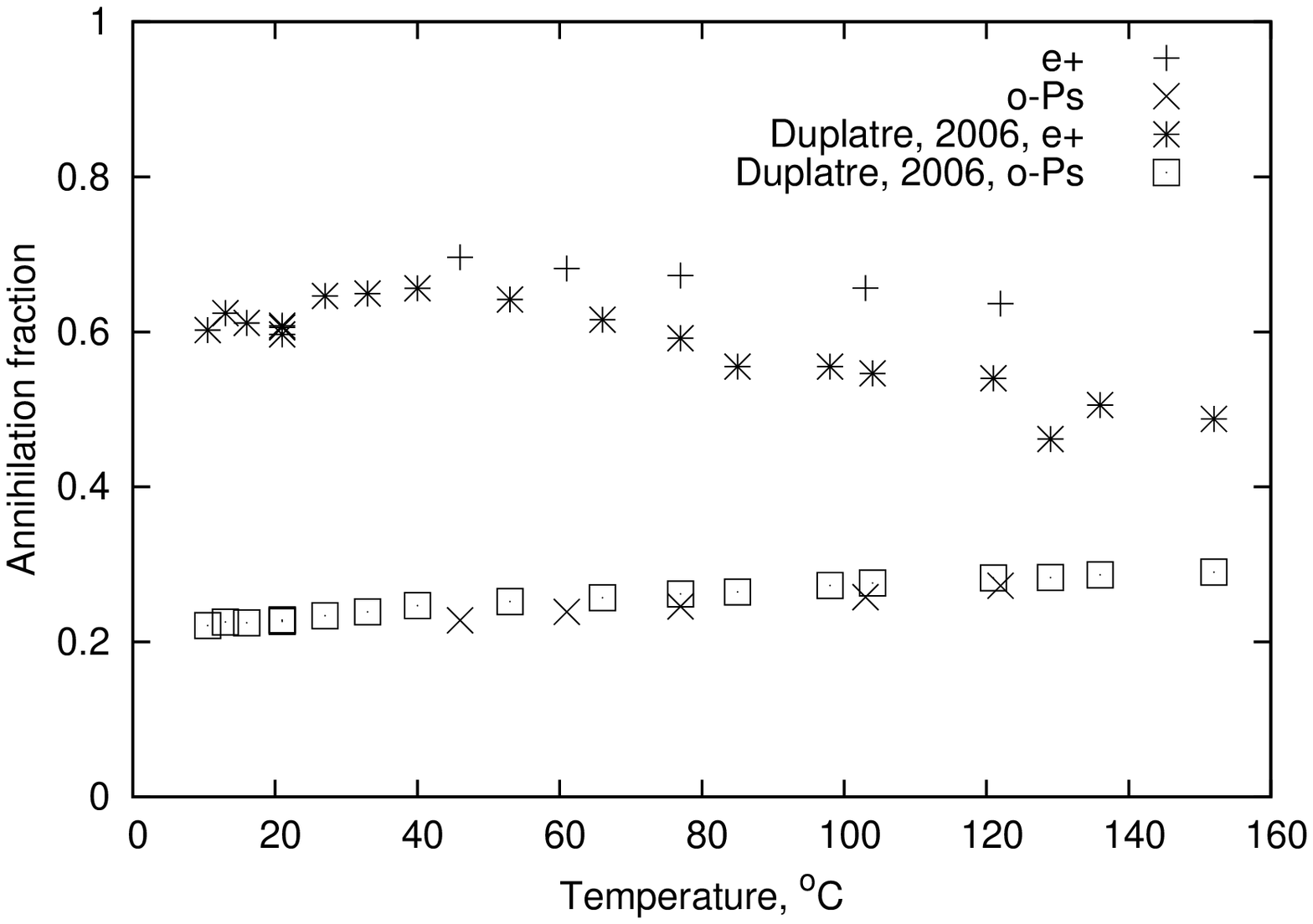} \includegraphics[scale=0.43]{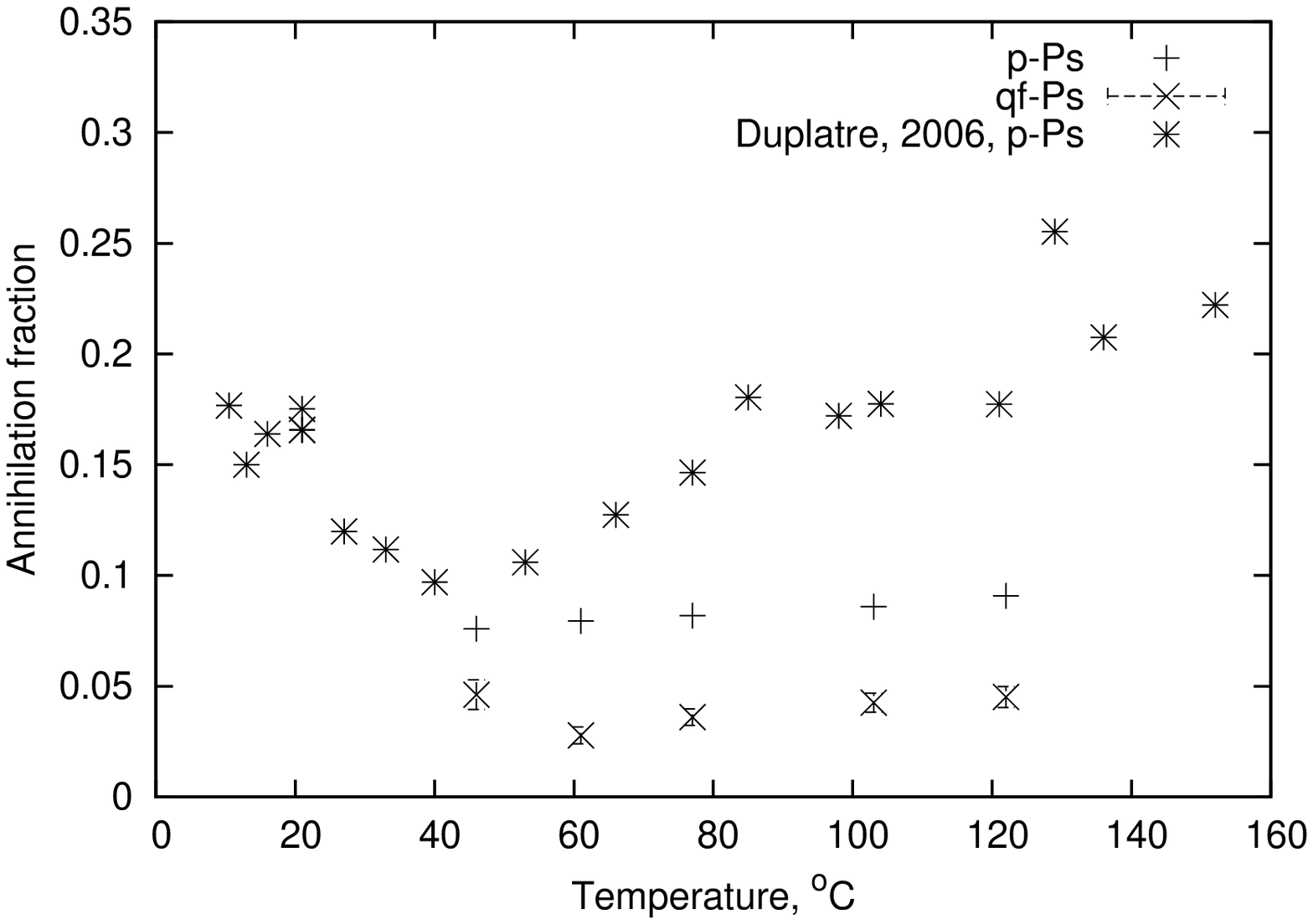}
 \caption{Temperature dependence of annihilation rates for glycerol}
 \label{fig:rate1}
\end{figure}

Annihilation rates (\ref{eq:h1})-(\ref{eq:h3}) drawn on Fig. \ref{fig:rate1} are compared to corresponding intensities from exponential decomposition. We observe a good agreement with reference data for o-Ps. The integral annihilation rate for p-Ps slightly differs from the estimates obtained from lifetime spectra. The estimates of errors for parameters, drawn on Fig. \ref{fig:rate1}, are calculated with standard error propagation formula and their value does not exceed the size of marker. Errors for adjustable parameters are calculated by minuit routine\cite{minuit}, which is used for optimization of collective chi-square function. 
%

\section{Conslusions}

With the specially developed theoretical model, we have described three experimental dependencies, which are Doppler and lifetime spectra, and $s(t)$ parameter. We have shown that it is necessary to account the presence of a delocalized Ps state at early stages of Ps interaction with the medium. In other words, the trapping rate $\kappa$ has finite value and the annihilation rate of qf-Ps differs from 0 outside the range of estimated errors.  
As a conclusion, we should note that this preliminary processing should not substitute the complimentary analysis of this data with radiolysis mechanisms taken into account. Such analysis is in the process of preparation by now.

\section{Acknowledgments}
The work of D.S. Zvezhinskiy and S.V. Stepanov is supported by Russian Fund for Basic Research (grant no. 11-03-01066).


\begin{thebibliography}{99}
\vspace{-2mm}
\bibitem{jop13} D.S. Zvezhinskiy, M. Butterling, A. Wagner, R. Krause-Rehberg and S.V. Stepanov // J. Phys.: Conf. Ser. Vol. 443 (2013) p 012057

\bibitem{Butt11} M. Butterling, W. Anwand, T.E. Cowan, A. Hartmann, M. Jungmann, R. Krause-Rehberg, A. Krille, A. Wagner // Nuclear Instruments and Methods in Physics Research B 269 (2011) 2623–2629

\bibitem{glyrad} J. Bonin, I. Lampre, P. Pernot, M. Mostafavi //  J. Phys. Chem. A 112 (2008) 1880-1886

\bibitem{Schneider} H. Schneider, A. Seeger, A. Siegle, H. Stoll, P. Castellaz, J. Major //  Applied Surface Science 116 (1997) 145-150

\bibitem{ryzhik} Gradshteyn and Ryzhik, Table of Integrals, Series, and Products // 4-th ed., Fizmatgiz, Moscow (1963)  (in Russian)

\bibitem{gly2009} Stepanov S., Zvezhinskii D., Duplatre G., Byakov V., Subrahmanyam V. // Mat.Sci. Forum. 2009. Vol. 607. Pp. 260–262

\bibitem{ste11} S.V. Stepanov, D.S. Zvezhinski, G. Duplâtre, V.M. Byakov, Yu.Yu. Batskikh, P.S. Stepanov // Materials Science Forum 666 (2011) 109-114

\bibitem{stebubble} K. V. Mikhin, S. V. Stepanov, and V. M. Byakov // High Energy Chemistry, Vol. 39, No. 1, 2005, pp. 36–43. Translated from Khimiya Vysokikh Energii, Vol. 39, No. 1, 2005, pp. 44–51

\bibitem{minuit} F. James // CERN Program Library Long Writeup D506: MINUIT Function Minimization and Error Analysis, Reference Manual, http://wwwasdoc.web.cern.ch/wwwasdoc/minuit/minmain.html

%
%
%
%
%

\end{thebibliography}
\end{document}